\documentclass{optica-article}

\journal{opticajournal} 

\articletype{Research Article}

\usepackage{lineno}
\usepackage{listings}
\usepackage[most]{tcolorbox}
\usepackage[breaklinks=true]{hyperref}

\newtcblisting[auto counter]{customlisting}[1][]{sharp corners, 
    colframe=gray,
    listing only, 
    listing options={basicstyle=\ttfamily,language=python},
}

\begin{document}

\newcommand{\code}[1]{\scalebox{0.9}[1.0]{\texttt{#1}}}
\newcommand{\urlnarrow}[1]{\scalebox{0.9}[1.0]{\url{#1}}}

\newcommand{\figref}[1]{Fig.~\ref{#1}}
\newcommand{\pfigref}[1]{Figure~\ref{#1}}
\newcommand{\secref}[1]{Sec.~\ref{#1}}

\newcommand{\sizeum}[2]{#1~$\mu$m~$\times$ #2~$\mu$m}

\newcommand{\nitride}[0]{Si\textsubscript{3}N\textsubscript{4}}
\newcommand{\oxide}[0]{SiO\textsubscript{2}}
\newcommand{\tio}[0]{TiO\textsubscript{2}}

\title{invrs-gym: a toolkit for nanophotonic inverse design research}
\author{Martin F. Schubert}
\address{invrs.io, Mountain View, CA 94041, USA}
\email{mfschubert@gmail.com}


\begin{abstract*} 
The \textit{invrs-gym} is a toolkit for research in nanophotonic inverse design, topology optimization, and AI-guided design. It includes a diverse set of challenges—representing a wide range of photonic design problems—with a common software interface that allows multiple problems to be addressed with a single code. The gym includes lightweight challenges enabling fast iteration as well as challenges involving design of realistic 3D structures, the solutions of which are suitable for fabrication. The gym is designed to be modular, enabling research in areas such as objective functions, design parameterizations, and optimization algorithms, and includes baselines against which new results can be compared. The aim is to accelerate the development and adoption of powerful methods for photonic design.

\end{abstract*}

\section{Introduction}
Inverse design has emerged as a powerful tool for challenging design problems in photonics and is routinely used in the research setting. Inverse design commonly involves large-scale optimization with thousands or millions of degrees of freedom along with constraints which ensure that solutions correspond to valid, manufacturable structures. Such problems are notoriously difficult, which has led to the application or development of a large number of solution techniques---including topology optimization \cite{elesin2014time, frellsen2016topology, Hammond:21, Hammond:22}, level set methods \cite{Vercruysse2019, piggott2017fabrication, piggott2020inverse}, schemes that optimize regular features such as holes or square pillars \cite{shen2015integrated, Tahersima2019, GOUDARZI2022105268}, machine learning methods \cite{Zhaocheng, Jiang2021, Wiecha:21}, and many others.

Works that introduce new solution techniques often demonstrate their performance with a novel design problem or a variant of some previously studied device. However, even small modifications to a given problem can alter the underlying optimization challenge, making comparisons to prior works difficult. For example, the problem difficulty is known to be dramatically affected by the target length scale of solutions and design region size \cite{Vercruysse2019, imageruler, Schubert2022}. Further, great care is required when developing a new implementation of a particular problem, so that even minutiae agree and to ensure convergence. Therefore, it would be beneficial for researchers to adopt a set of standard test problems with a reference implementation for which objective values and computational performance can be directly compared. 

Standardized computational problems fall under the common task method paradigm \cite{common_task_method}, which is used widely in machine learning \cite{deng2009imagenet, lecun1998gradient, krizhevsky2009learning} and many other fields from protein structure prediction \cite{posebusters, basu2016dockq} to the social sciences \cite{common_task_method_social_science}, although not in photonic design. However, the need for standard benchmarks for inverse design has previously been identified in \cite{Angeris:21, sigmund2022benchmarking} and major progress toward this goal was recently made in \cite{imageruler}, which defined a diverse set of nanophotonic design test problems, provided example solutions, and introduced \code{imageruler}---a software package for \textit{a posteriori} length scale evaluation. The goal of this work is a further step toward standardized, easily-used test problems and evaluation metrics that facilitate reproducible research and the development of new and powerful inverse design methods. To this end are the following contributions: (1) the \textit{invrs-gym}, a collection of photonic inverse design challenges with a common software interface suitable for use in an optimization context; and (2) the \textit{invrs-leaderboard}, a dataset of solutions to gym challenges, including those from prior works and new solutions from this work.

All gym challenges are based on problems from prior work; several are test problems specified in \cite{imageruler} and \cite{Schubert2022}, another is from the case study of a commercial optical design tool, and others are from actual design problems for imaging, large-area metasurfaces, and quantum information processing applications where researchers designed, fabricated, and characterized photonic components. In several cases, gym challenges represent the first implementation of the design problem using non-commercial tools, or the first that is readily used with modern software stacks that enable automatic differentiation and have become ubiquitous in machine learning research. Therefore, the gym should be highly relevant and accessible to photonics and inverse design researchers as well as the machine learning and broader optimization communities.

The gym and leaderboard are open source software and available at \href{https://github.com/invrs-io/gym}{github.com/invrs-io/gym} and \href{https://github.com/invrs-io/leaderboard}{github.com/invrs-io/leaderboard}. The software architecture and design choices are discussed in \secref{section:software_design}. \secref{section:challenges} reviews all gym challenge problems and gives example solutions. Finally, \secref{section:conclusion} gives concluding remarks.

\section{Software design}
\label{section:software_design}
\subsection{Gym}

\begin{figure}[b]
\centering\includegraphics{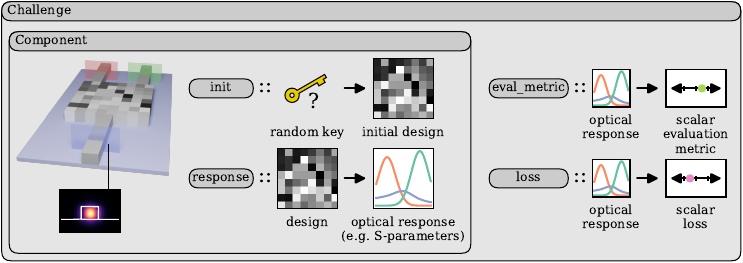}
\caption{The software architecture of the gym; classes are depicted as titled boxes and methods as \code{fn::a}$\rightarrow$\code{b}, which indicates a function taking type \code{a} as input and returning type \code{b}. The \code{Challenge} class has a \code{Component} attribute which represents the physical structure to be designed and details of the optical excitation. The \code{init} and \code{response} methods return initial design parameters and the (generally vector-valued) optical response of the component, respectively. The desired behavior of the component is encoded in methods of the \code{Challenge}. The \code{eval\_metric} computes a scalar value from the optical response, with higher values indicating more desirable performance; it is often only piecewise continuous. The \code{loss} function computes a scalar where smaller values correspond to better performance, and is generally better suited as an optimization objective.}
\label{figure:software}
\end{figure}

\pfigref{figure:software} depicts core elements of the software architecture for the gym. The primary types are the \code{Challenge} and \code{Component} objects. The \code{Component} represents the component to be designed---its physical structure and surroundings, and the optical excitation conditions such as excitation wavelength or incident mode (i.e. the waveguide mode or angle/polarization of a plane wave). Initial, random design parameters can be obtained from the \code{init} method. The \code{Component} also provides a \code{response} method, which takes design parameters as input, triggers a simulation, and returns the optical response (e.g. wavelength-dependent scattering parameters).

Each \code{Challenge} has a \code{Component} attribute as well as two methods that encode the desired performance of the component. The \code{eval\_metric} method computes the evaluation metric  $\mathscr{M}$ defined for each challenge---a measure of the quality of a given optical response, with higher values being better. Evaluation metrics for challenges are defined in various ways (\secref{section:challenges}), but in many cases have a worst-case form (e.g. the minimum efficiency across a range of wavelengths) which can be poor choice for direct use as an optimization objective. For this reason, the \code{Challenge} also has a \code{loss} method which is generally smoother, better suited to gradient-based optimization, but less fully reflects the response quality. Opposite to the evaluation metric, lower values of loss are better. The entire architecture supports automatic differentiation, so that gradients of loss with respect to design parameters can easily be computed. Basic usage of the gym is demonstrated in Appendix A.

The design parameters obtained from the \code{init} and \code{response} method represent the \textit{physical} design parameters. They include \text{density arrays}: arrays with values between 0 and 1 that are mapped to the permittivity at a particular point in the design region by interpolation \cite{CHRISTIANSEN201923}. In some cases, design parameters also include layer thicknesses. Valid solutions to a challenge must satisfy certain constraints, e.g. thicknesses must lie in some range, and density arrays must be binary and have a minimum length scale larger than a target value. These constraints are packaged together with the design parameters as metadata using types defined in the \code{totypes} package \cite{totypes-repo}.

Inverse design methods generally avoid the direct optimization of physical design parameters, and instead parameterize these with \textit{latent} variables \cite{elesin2014time, frellsen2016topology, Hammond:21, Hammond:22, Vercruysse2019, piggott2017fabrication, piggott2020inverse, shen2015integrated, Tahersima2019, GOUDARZI2022105268, Zhaocheng, Jiang2021, Wiecha:21, filter-project} e.g. to obtain solutions lacking non-manufacturable features. Thus, the physical design parameters exposed by the gym interface represent typically-intermediate quantities, and the gym itself does not constitute an end-to-end design framework as has been the objective of several recent works \cite{hazineh2022dflat, Yeung2023}. The gym eschews this goal in favor of a modular design that facilitates research. For example, the optical response for latent design variables of some arbitrary parameterization can be found by composing the parameterization with the \code{response} method. Various formulations of the scalar loss can can be explored using the challenge evaluation metric as an independent measure of solution quality. These can be compared to minimax optimization schemes that consider multiple objectives computed from the optical response. The effect of length scale constraints, initialization, and simulation settings (e.g. resolution) can also easily be explored with the modular gym design.

Besides enabling research, the software design of the gym is intended to support the development and adoption of an inverse design software ecosystem for widespread use. It is unlikely that any particular inverse design technique will be a good choice across the full range of photonic design challenges \cite{no_free_lunch}, and so practitioners should be free to swap algorithms and compare performance. This can be difficult with commercial or end-to-end design tools---where one is limited to the set of algorithms implemented by the chosen tool---and is an argument for a decoupled, modular software architecture. Since the physical parameters have relevance in all cases, regardless of parameterization, they are a natural choice for the interface of this architecture. In addition to this benefit, the narrower focus of the gym (and therefore, of complementary software ecosystem components) is expected to yield reduced software development and maintenance burden.

\subsection{Leaderboard}
The leaderboard is a relatively lightweight package that includes code to automatically evaluate solutions to gym challenges, existing solutions to the challenges (i.e. physical parameters), and a table of their evaluation results. The evaluation involves computing minimum length scale using \code{imageruler} and the challenge evaluation metric. The calculation of evaluation metric for the leaderboard relies on the gym, but differs from its calculation in an optimization setting in a few ways: while the gym code can be run on GPUs, leaderboard evaluations are forced to run on CPU with 64-bit precision. In addition, in some cases the number of wavelengths or simulation resolution are increased over their default values.

\section{Challenges}
\label{section:challenges}
\subsection{Metagrating}
\begin{figure}[b!]
\centering\includegraphics{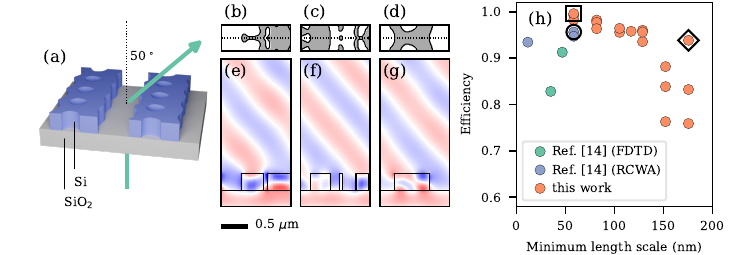}
\caption{(a) Schematic depiction of the metagrating challenge based on \cite{imageruler}. A 1050 nm $x$-polarized plane wave is incident from a \oxide{} substrate on a Si metasurface with air ambient. The objective is to couple light into the (+1, 0) diffraction order at a 50$^\circ$ angle. (b) A metagrating design from \cite{imageruler}. The design defines the structure within a \sizeum{0.525}{1.37} unit cell; gray regions are comprised of Si while white regions are areas where Si is etched away. The dotted line indicates an axis of symmetry. (c-d) Two metagrating designs from this work, with smaller and larger length scale. (e-g) The $x$-component of electric field in an $xz$ cross-section at the dotted line in (b-d). (h) The efficiency of beam deflector designs from \cite{imageruler} (generated with either an FDTD- or RCWA-backed topology optimization pipeline) and introduced in this work, as a function of the measured minimum length scale. The black circle, square, and diamond identify designs (b-d).}
\label{figure:metagrating}
\end{figure}

Beam deflectors are a common application for inverse design \cite{Sell2017, metagrating-01, Sabzevari:24, Ye:22, Chen:21, Xiao:24, Jenkins2023} and find use in fields such as augmented reality waveguides. The metagrating challenge is based on the \textit{3D metagrating} test problem from \cite{imageruler} and is schematically depicted in \figref{figure:metagrating} (a). It requires the design of a biperiodic patterned layer to deflect a normally-incident, monochromatic, $TM$-polarized (electric field in $x$ direction) plane wave into the (+1, 0) diffraction order at an angle of 50 degrees. Light with 1050 nm wavelength is incident from a silicon oxide substrate; the patterned layer is 0.325 $\mu$m thick and is composed of silicon with air filling the void regions.

The metagrating optimization variable is an array defining the pattern within the \sizeum{0.525}{1.37} unit cell, and its evaluation metric $\mathscr{M}$ is the polarization-summed diffraction efficiency for the target order, i.e.
\begin{equation}
\mathscr{M} = \left|S_{(+1, 0)}^{TM\rightarrow TM}\right|^2 + \left|S_{(+1, 0)}^{TM\rightarrow TE}\right|^2
\end{equation}
The gym implementation of the metagrating challenge uses FMMAX, an open-source RCWA code \cite{fmmax} that gives results in strong agreement with those reported in \cite{imageruler}  (Appendix B). Due to its low compute cost (Appendix C), small design region, and relatively simple objective, the metagrating challenge is a good initial choice for rapid evaluation of optimization methods.

Metagrating designs were created in \cite{imageruler} using topology optimization pipelines backed by Meep (FDTD) \cite{meep} and Reticolo (RCWA) \cite{hugonin2021reticolo}; an example design is in \figref{figure:metagrating} (b). Additional metagrating designs generated in this work are shown in \figref{figure:metagrating} (c-d). High efficiency beam deflection for the three designs is evident in \figref{figure:metagrating} (e-g), which plot the $x$-component of electric field for a plane wave incident from the substrate. The efficiency as a function of measured minimum length scale is shown in \figref{figure:metagrating} (h) for all designs from \cite{imageruler} and this work.

\subsection{Diffractive splitter}
\begin{figure}[b!]
\centering\includegraphics{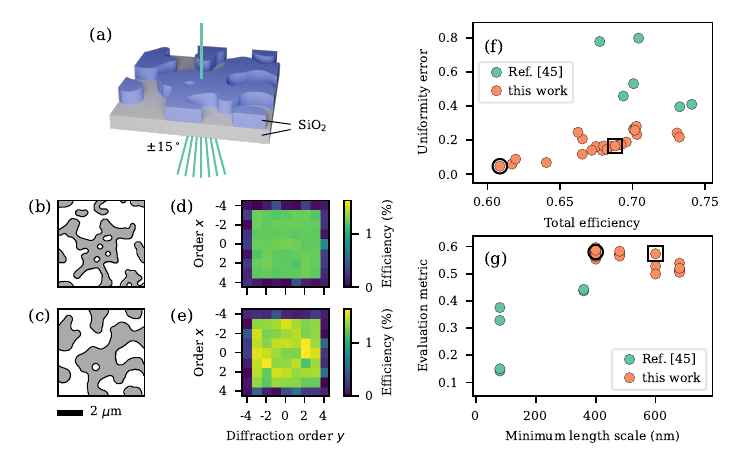}
\caption{(a) Schematic depiction of the diffractive splitter challenge based on \cite{LightTrans} A 632.8 nm y-polarized plane wave is incident from air upon a patterned \oxide{} layer on an \oxide{} substrate. The objective is to split light evenly into 49 beams, i.e. transmitted diffraction orders from $(-3, -3)$ to $(+3, +3)$. (b-c) Two diffractive splitter designs with differing minimum length scale. Gray regions are composed of \oxide{} and white regions are areas where \oxide{} should be etched away. (c-d) Efficiency of diffraction into transmitted orders for the two splitter designs. (e) The uniformity error as a function of total efficiency for designs from \cite{LightTrans} and this work. The black circle and square identify designs (b) and (c). (f) The evaluation metric (defined as the product of total efficiency and (1 - uniformity error)) as a function of the measured minimum length scale.}
\label{figure:diffractive_splitter}
\end{figure}

Diffractive optical splitter design is a common problem in photonics with applications including structured light projection \cite{structured_light_1, structured_light_2} and multifocal microscopy \cite{multifocal_1, multifocal_2}. The diffractive splitter challenge is based on \cite{LightTrans} and is schematically depicted in \figref{figure:diffractive_splitter} (a). A $TE$-polarized plane wave at 632.8 nm is normally incident from air upon a biperiodic patterned \oxide{} layer on an \oxide{} substrate. The challenge objective is to split power equally into transmitted diffraction orders $(-3, -3),\ldots,(+3, +3)$, corresponding to angles of $\pm15^\circ$ for the \sizeum{7.2}{7.2} unit cell dimensions. The diffractive splitter optimization variable is an array defining the pattern and the scalar thickness of the patterned layer. Its evaluation metric is defined in terms of the polarization summed diffraction efficiency $\eta_{(i,j)}=|S_{(i, j)}^{TE\rightarrow TE}|^2 + |S_{(i, j)}^{TE\rightarrow TM}|^2$ for target order $(i, j)$,
\begin{equation}
\mathscr{M} = \left(\sum_{i,j} \eta_{(i, j)} \right)
\left(1 -
\frac{\max_{i,j} \eta_{(i,j)} - \min_{i,j} \eta_{(i,j)}}{\max_{i,j} \eta_{(i,j)} + \min_{i,j} \eta_{(i,j)}}
\right)
\label{eq:diffractive_splitter}
\end{equation}
where the first term is the total efficiency, and the fraction in the second term is the uniformity error.

Diffractive splitter designs in \cite{LightTrans} were parameterized by a grid of square pillars and optimized with a hybrid optimization strategy to restrict run times. Initial steps using the iterative Fourier transform algorithm and thin element approximation, after which refinement and final evaluation were done using an RCWA model. A similar scheme was used in \cite{Kim:20}, except that a filter-and-projection parameterization \cite{filter-project} was used. The diffractive splitter challenge is implemented using FMMAX and achieves sub-second effective step times (Appendix A) and excellent agreement with RCWA results from \cite{LightTrans} (Appendix B), enabling the full optimization to be carried out with an accurate model. \figref{figure:diffractive_splitter} (b-c) depict two designs generated for this work with 400 nm and 600 nm minimum feature size, respectively. The diffraction efficiency into each order is shown in \figref{figure:diffractive_splitter} (d-e), with design (b) showing particularly low nonuniformity error but with slightly lower efficiency than design (c).

The uniformity error is plotted as a function of efficiency in \figref{figure:diffractive_splitter} (f) for all designs from \cite{LightTrans} and this work. The results suggest a tradeoff exists between the two quantities, making the diffractive splitter a problem for research on multiobjective optimization or scalarization strategies. The evaluation metric is plotted as a function of measured minimum length scale in \figref{figure:diffractive_splitter} (g). Small length scale values for designs from \cite{LightTrans} are due to the presence of specific features that arise from the square pillar parameterization, as discussed in Appendix D.

\subsection{Meta-atom library}
Large-area metasurfaces built from libraries of meta-atoms are an emerging platform for compact optical devices with tailored functionality in applications such as imaging, sensing, and displays \cite{pestourie2018inverse, li2022inverse, Chen2023, phan2019high, zhou2024large}. The meta-atom library challenge is based on \cite{Chen2023} and entails the design of eight dispersion-engineered meta-atoms for use in metasurfaces that operate across the visible spectrum. Meta-atoms lie within a square 400 nm unit cell and are comprised of \tio{} structures lying on an \oxide{} substrate. The optimization variables consist of arrays defining the meta-atom patterns and a scalar giving the \tio{} height. Patterns must be free of \tio{} at the border to ensure that meta-atoms can be arbitrarily tiled without introducing non-manufacturable features, and should have reflection symmetry along the $x$ or $y$ axis to achieve polarization-independent response \cite{Chen2023}.

The optical response of the meta-atom library consists of the per-atom wavelength-dependent zeroth-order complex transmission coefficients $(t_n^{j,R\rightarrow R}, t_n^{j,R\rightarrow L}, t_n^{j,L\rightarrow R}, t_n^{j,L\rightarrow L})$, where $n$ indexes the meta-atom, $j$ indexes the wavelength, and e.g. $R\rightarrow L$ indicates right-hand circularly-polarized incident light converted to left-hand circular polarization. When meta-atoms have the appropriate symmetry, $t_n^{j,R\rightarrow R} = t_n^{j,L\rightarrow L}$ and $t_n^{j,R\rightarrow L} = t_n^{j,L\rightarrow R}$. From the transmission coefficients, the fields resulting from plane wave excitation of an arbitrary arrangement of meta-atoms can be approximated by spatially stitching the per-atom transmission coefficients \cite{pestourie2018inverse}. For the evaluation metric, the meta-atom arrangement implementing a 1D grating is considered; its diffraction into various orders is approximated by the discrete Fourier transform of the stitched amplitudes, i.e.
\begin{equation}
S_{(k, 0)}^{j,A\rightarrow B} = 
\frac{1}{8}
\sqrt{\frac{n_{substrate}}{n_{ambient}}}
\sum_{n=0}^{7}
t_n^{j,A\rightarrow B} \ e^{-i \pi k n / 4}
\end{equation}
where $n_{substrate}$ and $n_{ambient}$ are the substrate and ambient refractive indices. Defining the polarization-summed diffraction efficiency as
\begin{equation}
\eta_{(k, 0)}^j = \left|S_{(k, 0)}^{j,R\rightarrow R}\right|^2 + \left|S_{(k, 0)}^{j,R\rightarrow L}\right|^2
\label{eq:efficiency}
\end{equation}
the evaluation metric for the meta-atom library challenge is the minimum relative efficiency across all wavelengths, i.e.
\begin{equation}
\mathscr{M} = \min_j \frac{\eta_{(+1, 0)}^j}{\sum_k \eta_{(k, 0)}^j}
\end{equation}
Note that due to the approximation involved in computing the relative efficiency, a higher evaluation metric does not guarantee greater relative efficiency for a metagrating built from the meta-atom library. For evaluation, the wavelengths are between 450 nm and 650 nm with 20 nm spacing. Due to its broadband performance objective and multiple design regions, the meta-atom library challenge represents a relatively advanced problem.

\begin{figure}[t]
\centering\includegraphics{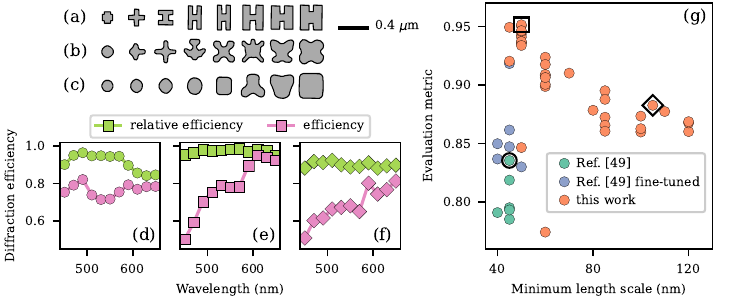}
\caption{Dispersion-engineered meta-atom libraries (a) obtained by particle swarm optimization in \cite{Chen2023} and (b-c) by topology optimization in this work. The meta-atoms are comprised of \tio{} and positioned on a \oxide{} substrate. The ordering of meta-atoms in (a-c) implements a 1D diffraction grating. (d-f) The diffraction efficiency and relative diffraction efficiency for designs (a-c), computed by discrete Fourier transform of the the stitched per-atom complex transmission coefficients. (g) The evaluation metric for all designs from the leaderboard dataset as a function of measured minimum length scale; the evaluation metric is found by taking the minimum relative efficiency across all wavelengths in (d-f). The circle, square, and diamond identify designs (a-c).}
\label{figure:meta_atom_library}
\end{figure}

Several meta-atom libraries were created in \cite{Chen2023} using a commercial FDTD solver and particle swarm optimization, with an example library shown in \figref{figure:meta_atom_library} (a). The gym meta-atom library challenge is backed by FMMAX; comparison of conserved and converted polarization phases shows the two methods to be in excellent agreement (Appendix B). Additional meta-atom libraries were created for this work by two methods: in the first method, designs from \cite{Chen2023} were fine-tuned using a level-set method. The second method used topology optimization, random initialization, and length scale constraints from 50 nm to 120 nm. \figref{figure:meta_atom_library} (b) and (c) show designs generated by toplogy optimization. The wavelength-dependent diffraction efficiencies and relative diffraction efficiencies (computed by Eq. \ref{eq:efficiency}) for designs (a-c) are plotted in \figref{figure:meta_atom_library} (d-f). The topology-optimized designs exhibit greater wavelength dependence for efficiency paired with lower wavelength dependence for relative efficiency, suggesting a tradeoff between these quantities. The evaluation metric of all designs from the dataset is plotted against the measured minimum length scale in \figref{figure:meta_atom_library} (d). 

\subsection{Bayer sorter}
\begin{figure}[t]
\centering\includegraphics{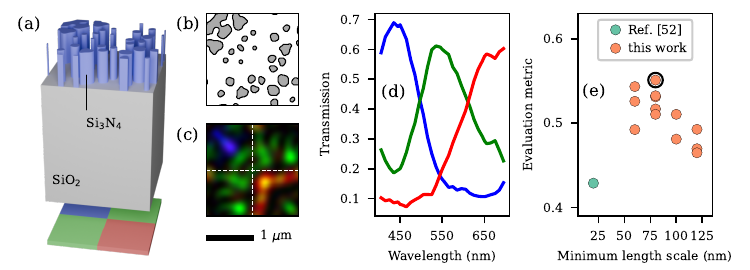}
\caption{(a) Schematic depiction of the Bayer color sorter challenge based on \cite{Zou2022}. Unpolarized visible light is normally incident from air upon a \nitride{} metasurface on \oxide{} spacer; the objective is to direct blue wavelengths into the blue subpixel, green wavelengths to the two green subpixels, and red wavelengths to the red subpixel. (b) An example Bayer color sorter design; gray regions are composed of \nitride{} and white regions are areas where the \nitride{} should be etched away. (c) Normalized $z$-oriented Poynting flux below the spacer for the design in (b). In (c), the red, green, and blue color channels are taken from the Poynting flux at 650 nm, 550 nm, and 450 nm wavelengths, and each is scaled to have a peak value of 1. (d) Transmission into the red, green, and blue subpixels as a function of wavelength. (e) Evaluation metric for Bayer color sorter designs as a function of the measured minimum length scale from \cite{Zou2022} and introduced in this work. The black circle identifies the design in (b).}
\label{figure:bayer}
\end{figure}

The Bayer sorter challenge is depicted in \figref{figure:bayer}; it entails the design of a biperiodic metasurface that focuses unpolarized broadband light in a wavelength-dependent manner upon an array of pixels arranged in a Bayer pattern. The metasurface and spacer are composed of \nitride{} and \oxide{} respectively, the pitch is 2 $\mu$m, and the layer thicknesses are additional optimization variables. Materials and dimensions are identical to the Bayer color sorter studied in \cite{Zou2022}. Similar color-sorting metasurfaces have been widely studied \cite{Peng:23, Miyata:21, Chen2017}, including metasurfaces for submicron pixels \cite{submicron_bayer, CatrysseZhaoJinFan, freeform_submicron}, RGB/NIR color sorting metasurfaces \cite{VIS/NIR1, VIS/NIR2, VIS/NIR3, VIS/NIR4}, and hybrid color/polarization-sorting metasurfaces \cite{ZOU2023107472}.

Bayer sorter designs were generated in \cite{Zou2022} using a genetic algorithm with a commercial FDTD solver; in the Bayer sorter challenge an FMMAX-backed model is used, which is in excellent agreement with the commercial solver (Appendix B). In this work, several sorter designs with varying length scale were generated. \figref{figure:bayer} (b) shows an example design with 80 nm minimum length scale. \figref{figure:bayer} (b) plots the normalized, polarization-averaged z-component of Poynting flux below the \oxide{} spacer for 450 nm, 550 nm, and 650 nm light. \figref{figure:bayer} (d) plots the transmission into subpixels of the three colors as a function of wavelength, showing the strong color sorting performance.

The Bayer sorter evaluation metric is,
\begin{equation}
\mathscr{M} = \min\left\{\left<T_B(\lambda_B)\right>, \left<T_G(\lambda_G)\right>, \left<T_R(\lambda_R)\right> \right\}
\end{equation}
where $\left<T_c(\lambda_c)\right>$ is the average transmission of wavelengths $\lambda_c$ associated with color $c$ into the subpixel(s) for color $c$. The test wavelengths are 433 nm and 467 nm for blue $(B)$, 533 and 567 nm for green $(G)$, and 633 and 667 nm for red $(R)$. The evaluation metric is plotted as a function of the measured minimum length scale in \figref{figure:bayer} (e) for designs introduced in this work and the design from \cite{Zou2022}. The small length scale values for designs from \cite{Zou2022} are due to the parameterization with square pillars (Appendix E).

\subsection{Metalens}
The metalens challenge is based on the \textit{RGB metalens} test problem from \cite{imageruler} and entails the design of a 10 $\mu$m wide, 1 $\mu$m tall one-dimensional lens that focuses 450 nm, 550 nm, and 650 nm normally-incident plane waves with in-plane electric field to a spot 2.4 $\mu$m above the metalens surface. Light is incident from a substrate with $n=2.4$, and solid and void regions of the lens have refractive index 2.4 and 1.0, respectively. Two metalens designs are depicted in \figref{table:metalens} (a-b) along with the electric field intensity, showing the desired focusing behavior. Similar structures were studied in \cite{Christiansen:21}, and the general problem of metalens design to achieve a single focus at multiple wavelengths is an important and widely studied problem \cite{Li2022, metalens2, metalens3, Christiansen:20, Chung:20}.

\begin{figure}[t]
\centering\includegraphics{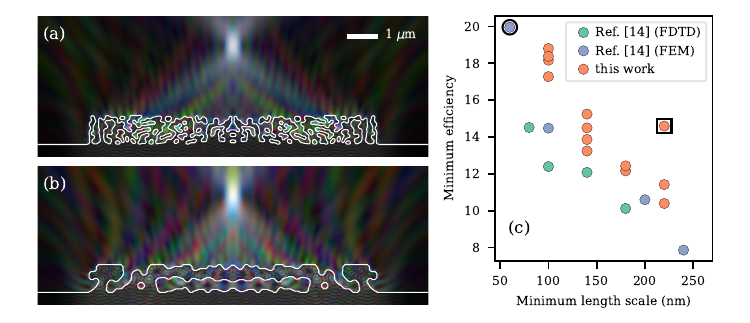}
\caption{(a) Electric field amplitudes and structure of a metalens challenge solution from \cite{imageruler}. Light with 450 nm, 550 nm, and 650 nm wavelengths having $x$-polarized (in-plane) electric field is incident from an $n=2.4$ substrate upon a 10$\times$1 $\mu$m metalens comprised of $n=2.4$ material with $n=1$ ambient. The blue, green, and red channels of the image are the electric field amplitudes for 450 nm, 550 nm, and 650 nm wavelengths, each scaled by the same factor so that a colorless pixel indicates that all wavelengths have equal amplitude at the pixel location. (b) A metalens design introduced in this work. (c) Evaluation metric for metalens designs as a function of the measured minimum length scale, including designs from FEM-backed and FDTD-backed topology optimization pipelines from \cite{imageruler} and new designs introduced in this work. The black circle and square identify designs (a) and (b).}
\label{figure:metalens}
\end{figure}

The metalens evaluation metric is the minimum intensity enhancement among the three wavelengths, i.e.
\begin{equation}
\mathscr{M} = \min_j \frac{\left|\mathbf{E}_j(\mathbf{r}_0)\right|^2 }{\left|\mathbf{E}^0_j(\mathbf{r}_0)\right|^2}
\end{equation}
where $\mathbf{E}_j$ is the electric field for wavelength $\lambda_j$ of the structure including metalens, $\mathbf{E}_j^0$ is the electric field of a structure without metalens, and $\mathbf{r}_0$ is the desired focus position, i.e. 2.4 $\mu$m above the center of the lens.

While \cite{imageruler} used FDTD- and FEM-based simulations to evaluate designs, the metalens challenge is implemented using FMMAX. This is done by slicing the 1 $\mu$m tall structure into layers having 40 nm thickness. Despite the relatively large number of layers, the compute cost remains reasonable due to the one-dimensional lens geometry. To validate the FMMAX implementation of the metalens challenge, all designs from \cite{imageruler} were re-simulated; results were found to be in excellent agreement with with those from FDTD and FEM (Appendix B).

The metalens challenge is one with several existing solutions across a wide range of length scales, making it a good choice when evaluating new inverse design schemes. In \cite{imageruler}, filter-and-projection topology optimization was used to generate designs with length scales between 70 and 250 nm. One example is the design in \figref{figure:metalens} (a), which was generated using the FEM-based simulation. For this work, additional designs were generated by topology optimization using the FMMAX-backed metalens challenge implementation; an example is in \figref{figure:metalens} (b). The evaluation metric of all designs is plotted as a function of the measured minimum length scale in \figref{figure:metalens} (c), and shows a decrease in the evaluation metric as the length scale increases.


\subsection{Ceviche challenges}
\begin{figure}[t]
\centering\includegraphics{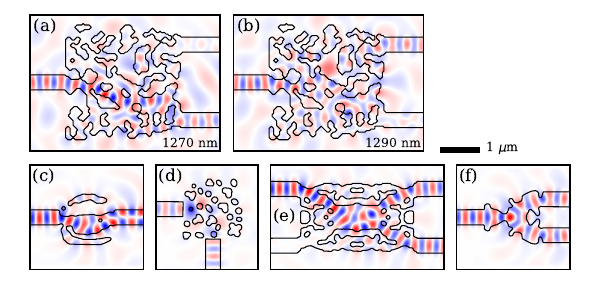}
\caption{Solutions to several Ceviche integrated photonic design challenges. (a-b) A solution to the lightweight wavelength demultiplexer challenge from \cite{surco} optimized using the SurCo strategy. Lightweight challenge simulations are less computationally expensive, involving coarser grid and fewer wavelengths. (c) A mode converter solution from \cite{imageruler} obtained using density-based topology optimization. (d) A waveguide bend from \cite{padhy2024PhoTOS} optimized using the PhoTOS scheme. (e) A beam splitter from \cite{Schubert2022} optimized using a conditional generator and gradient estimator strategy. (f) A power splitter optimized using a levelset parameterization.   Depicted fields for (c-f) are at 1280 nm, and at 1270 nm and 1290 nm for (a) and (b). All structures are excited from the left with the fundamental waveguide mode.}
\label{figure:Ceviche}
\end{figure}

The Ceviche challenges are integrated photonics test problems introduced in \cite{Schubert2022} and open-sourced in \cite{ceviche-challenges}. These include ten variants depicted in \figref{figure:Ceviche}---requiring the design of a beam splitter, mode converter, power splitter, waveguide bend, and a two-channel wavelength demultiplexer, with standard and "lightweight" versions of each. Lightweight challenges trade off accuracy for lower computational cost and are particularly suited as vehicles for algorithm development. The challenges are named for their use of the Ceviche two-dimensional finite-difference frequency domain (FDFD) simulation engine \cite{Hughes2019}, and feature design regions from \sizeum{1.6}{1.6} to \sizeum{6.4}{6.4} in size connected to waveguides having 400 nm width. The permittivity of core and cladding material are that of Si and \oxide{}, respectively, and operation is in the 1260--1300 nm wavelength range.

Ceviche challenge components have 2--4 waveguides connected to the design region; all are excited from the top left waveguide (with index 1). The optical response consists of wavelength-dependent scattering parameters $S_{(i,1)}^j$, where $i$ indexes the waveguides and $j$ the wavelengths. The power transmission coefficients $|S_{(i,1)}^j|^2$ define a multi-dimensional space; each Ceviche challenge defines a target orthotope within the space, i.e. lower bound $l_i^j$ and upper bound $u_i^j$ for each transmission coefficient. The evaluation metric is defined as,
\begin{equation}
\mathscr{M} = \min_{i,j} \frac{1}{u_i^j - l_i^j}
\begin{cases}
|S_{(i,1)}^j|^2 - l_i^j & u_i^j = 1 \\[6pt]
u_i^j - |S_{(i,1)}^j|^2 & l_i^j = 0 \\[6pt]
\min \left\{|S_{(i,1)}^j|^2 - l_i^j, \ u_i^j - |S_{(i,1)}^j|^2\right\} & \text{otherwise}
\end{cases}
\end{equation}
which is positive when transmission is inside the target orthotope. Here, negative $|S_{(i,1)}^j|^2 - l_i^j$ constitutes a violation of the lower bound for power transmission into mode $i$ for wavelength $j$, and negative $u_i^j - |S_{(i,1)}^j|^2$ is a violation of the upper bound. The piecewise definition is used so that the upper and lower bounds are disregarded when they match the theoretical extrema, i.e. 0 and 1, respectively.

Following their introduction in \cite{Schubert2022}, the Ceviche challenges have been adopted as test problems in several other works. Lightweight challenges including the wavelength demultiplexer were studied in \cite{surco, ferber2023genco} using the SurCo scheme for nonlinear combinatorial optimization problems. A demultiplexer design from \cite{surco} is shown in \figref{figure:Ceviche} (a-b). The mode converter problem was adopted as a test problem in \cite{imageruler}, and a filter-and-projection \cite{filter-project} topology optimization scheme was compared to the conditional generator and gradient estimator approach of \cite{Schubert2022}. An example filter-and-projection design is shown in \figref{figure:Ceviche} (c). Mode converters and waveguide bends were studied in \cite{padhy2024PhoTOS} using the PhoTOS scheme, in which solutions are parameterized by shapes from a predefined library that are translated, scaled, and oriented while respecting length scale constraints. An example PhoTOS design is shown in \figref{figure:Ceviche} (d). The remaining challenges are illustrated in \figref{figure:Ceviche} (e) and (f): a beam splitter design from \cite{Schubert2022}, and a power splitter from this work optimized using a level set parameterization \cite{Vercruysse2019}. The scattering parameters for all designs are given in Appendix E.

Since the Ceviche challenges involve two-dimensional simulations, they constitute a test suite that is not fully representative of actual integrated photonic design challenges. However, methods that were developed using the Ceviche challenges have been shown capable of producing high-performance solutions for realistic design challenges \cite{Cheung:24}, thereby establishing the utility of the suite.

\subsection{Photon extractor}
\begin{figure}[t]
\centering\includegraphics{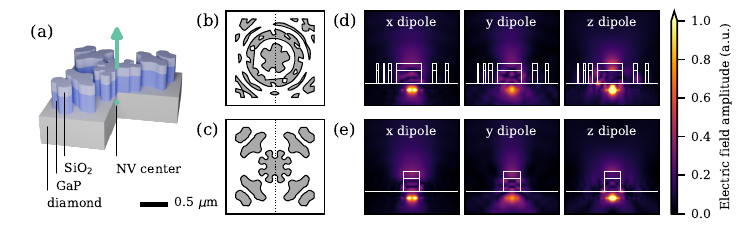}
\caption{(a) Schematic depiction of the photon extractor challenge based on \cite{Chakravarthi:20}. A NV center emitting at 637 nm is positioned 100 nm below the surface of a diamond substrate; the objective is to maximize light collected above the defect. The challenge entails design of a GaP metastructure with 0.25 $\mu$m thickness that is capped by a 0.13 $\mu$m oxide patterning hard mask. (b) A photon extractor design from \cite{Chakravarthi:20} generated using a topology optimization pipeline with custom finite-difference frequency-domain solver. Gray regions represent GaP, and white regions are areas where GaP is to be etched away. (c) A design generated using the FMMAX-backed photon extractor challenge with relatively larger minimum length scale. (d-e) Electric field amplitudes for $x$-, $y$-, and $z$-oriented dipoles in an $xz$ cross-section for the designs (b-c).}
\label{figure:photon_extractor}
\end{figure}

The photon extractor challenge is based on \cite{Chakravarthi:20} and depicted in \figref{figure:photon_extractor} (a). A Nitrogen vacancy (NV) center is positioned 100 nm below the surface of a diamond substrate, above which is a patterned 250 nm GaP layer capped with 130 nm \oxide{} used as a patterning hard mask. The design variable is the array defining the GaP pattern within a \sizeum{1.5}{1.5} region centered over the NV center. The objective is to maximize the collected power for photons at the 637 nm zero-phonon line, as needed for quantum information applications \cite{Chakravarthi:20}. Maximizing collected power and involves both extraction efficiency and local density of states (LDOS) enhancement; LDOS enhancement has been studied in several previous works \cite{wang2018maximizing, icsiklar2022trade, miller2016fundamental, liang2013formulation} and is a useful ingredient for a test problem due to the typically high sensitivity of enhancement to length scale \cite{imageruler}.

Emission by the NV center is modeled by separately considering $x$-, $y$-, and $z$-oriented dipoles. The optical response of the photon extractor consists of the dipole-orientation-dependent total emitted power, total extracted power, and total collected power, with the collected power defined as the total power passing through a \sizeum{1.5}{1.5} monitor positioned 0.4 $\mu$m above the oxide. The response also includes the powers for a reference structure consisting of the bare diamond substrate, which can be used to calculate flux and LDOS enhancements. The evaluation metric is the dipole-orientation-averaged enhancement in collected flux, i.e.
\begin{equation}
\mathscr{M} = \frac{\Phi_{x, collected} + \Phi_{y, collected} + \Phi_{z, collected}}
{\Phi_{x, collected}^0 + \Phi_{y, collected}^0 + \Phi_{z, collected}^0}
\end{equation}

In \cite{Chakravarthi:20}, designs were generated using an FDFD-backed topology optimization pipeline and subsequently evaluated using FDTD. The photon extractor model is implemented with FMMAX with perfectly matched layer boundary conditions; this has computational efficiency benefits for modeling spontaneous emission \cite{fmmax} but is inherently limited in the modeling of point dipoles due to the use of a Fourier basis. Nevertheless, LDOS and flux enhancements computed by FMMAX approach the values reported in \cite{Chakravarthi:20} as the number of Fourier orders increases (Appendix B).

A photon extractor design from \cite{Chakravarthi:20} is shown in \figref{figure:photon_extractor} (b); its measured minimum length scale is below 20 nm due to the presence of sharp features (Appendix D). A design generated in this work by topology optimization with 80 nm minimum length scale constraints is shown in \figref{figure:photon_extractor} (c). Electric field cross-sections in the $xz$ plane for by $x$-, $y$-, and $z$-oriented dipoles for the two designs are shown in \figref{figure:photon_extractor} (d-e). The fields in all cases are comparable, and the two designs achieve similar evaluation metrics of 11.3 and 12.1, respectively.

\section{Concluding remarks}
\label{section:conclusion}
This work introduces the \textit{invrs-gym} and \textit{invrs-leaderboard}, which aim to accelerate the development and adoption of powerful methods for photonic design. The gym provides modern, open-source implementations of design challenges that have been previously identified as useful test problems or as relevant design problems in fields such as large-area metasurfaces or quantum information processing. In many cases, gym implementations of these design challenges are the first using non-commercial tools, or the first using a modern software framework that supports automatic differentiation. Each challenge is validated by detailed comparisons to prior work. Gym challenges are directly usable in an optimization setting, and are exercised in this work to generate new solutions for several challenges. These, along with solutions from prior works, are included in the leaderboard---a dataset of solutions to gym challenges with a mechanism for new, verified contributions. It is hoped that researchers of nanophotonic inverse design find the gym approachable, and that it reduces the effort required develop and evaluate new methods.

In the future, the gym could be extended to include new challenges. In particular, problems that are found to present difficulties (which differ categorically from those of existing challenges) to design algorithms would be of value. It would also be useful to augment the leaderboard with solutions generated using other design methodologies, so that it becomes a resource that aids designers in selecting optimization schemes for individual nanophotonic design problems. Finally, it would be valuable to undertake a comparison of design methods that considers more than just objective value, e.g. computational efficiency or the robustness of solutions to fabrication variations, as these represent additional important considerations in the selection of an optimization method.

\pagebreak

\section*{Appendix A: Example gym usage in optimization setting}

The following Python code exercises gradient descent optimization of the metagrating challenge.

\begin{customlisting}
import jax
from invrs_gym import challenges

challenge = challenges.metagrating()
params = challenge.component.init(jax.random.PRNGKey(0))
learning_rate = 0.1

def loss_fn(params):
  response, _ = challenge.component.response(params)
  return challenge.loss(response), response

for _ in range(20):
  grad, response = jax.grad(loss_fn, has_aux=True)(params)
  params = jax.tree_util.tree_map(
    lambda p, g: p - learning_rate * g,
    params,
    grad,
  )

print(f"eval_metric={challenge.eval_metric(response)}")
\end{customlisting}

\pagebreak
\section*{Appendix B: Validation of simulation models}
Table \ref{table:metagrating_comparison} compares the reported diffraction efficiency of five metagrating designs from \cite{imageruler} to the efficiency calculated with the FMMAX-backed implementation of the metagrating challenge. In \cite{imageruler}, values are reported for both FDTD and RCWA models of the metagrating; the efficiency values for all solvers are in close agreement.

\begin{table}[b]
\footnotesize
\begin{center}
\begin{tabular}{ c | c | c | c }
  & Reticolo (RCWA) \cite{imageruler} & Meep (FDTD) \cite{imageruler} & FMMAX (RCWA) \\ 
 \hline
 device 1 & 0.957 & 0.955 & 0.956 \\  
 device 2 & 0.933 & 0.938 & 0.935 \\  
 device 3 & 0.966 & 0.950 & 0.947 \\  
 device 4 & 0.933 & 0.925 & 0.913 \\  
 device 5 & 0.841 & 0.843 & 0.828
\end{tabular}
\end{center}
\caption{Diffraction efficiency for metagrating designs as reported in \cite{imageruler} and computed using the FMMAX-backed implementation of the metagrating challenge. Devices 1-3 were generated using an RCWA-backed topology optimization pipeline; devices 4-5 were generated using an FDTD-backed pipeline.}
\label{table:metagrating_comparison}
\end{table}

Table \ref{table:diffractive_splitter_comparison} compares the total efficiency and uniformity error for six diffractive splitter solutions from \cite{LightTrans}, as reported in \cite{LightTrans} and calculated with the FMMAX-backed implementation of the diffractive splitter challenge. Values in \cite{LightTrans} were computed using the LightTrans RCWA package and are in close agreement with those obtained using FMMAX.

\begin{table}[b]
\footnotesize
\begin{center}
\setlength{\tabcolsep}{4pt}
\begin{tabular}{ c | c c | c c }
  & \multicolumn{2}{c}{LightTrans (RCWA) \cite{LightTrans}} &
  \multicolumn{2}{c}{FMMAX (RCWA)} \\ 
 \hline
  & total efficiency & uniformity error & total efficiency & uniformity error \\
 \hline
device 1 & 0.705 & 0.828 & 0.704 & 0.798 \\
device 2 & 0.702 & 0.568 & 0.701 & 0.531 \\
device 3 & 0.738 & 0.424 & 0.741 & 0.410 \\
device 4 & 0.679 & 0.806 & 0.677 & 0.778 \\
device 5 & 0.694 & 0.470 & 0.694 & 0.459 \\
device 6 & 0.729 & 0.429 & 0.732 & 0.396 \\
\end{tabular}
\end{center}
\caption{Figures of merit for diffractive splitter designs as reported in \cite{LightTrans} and computed using the FMMAX-backed implementation of the diffractive splitter challenge. The designs in \cite{LightTrans} were generated using the thin element approximation method, and evaluated using the LightTrans RCWA package. Designs 4-6 are identical to 1-3 except that layer thickness was separately optimized as a post-processing step.}
\label{table:diffractive_splitter_comparison}
\end{table}

\begin{figure}[t]
\centering\includegraphics{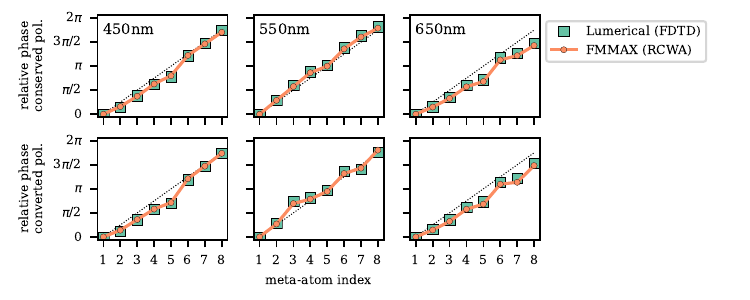}
\caption{Comparison of relative phase for conserved and converted polarizations for meta-atoms from \cite{Chen2023}, as computed by Lumerical FDTD in \cite{Chen2023} and using the FMMAX-backed implementation of the meta-atom library challenge. The dotted line indicates the ideal phase covering a range of $2\pi$.}
\label{figure:meta_atom_library_validation}
\end{figure}

\pfigref{figure:meta_atom_library_validation} plots the relative phase for conserved and converted polarizations for the eight meta-atoms as shown in Figure 2 (a-b) of \cite{Chen2023} and as computed using FMMAX. The values from \cite{Chen2023} were computed using Lumerical FDTD and are in close agreement with the FMMAX values, validating the implementation of the meta-atom library challenge.

\begin{figure}[t]
\centering\includegraphics{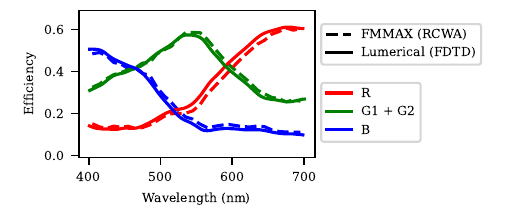}
\caption{Comparison of wavelength-dependent efficiency as computed by Lumerical FDTD in \cite{Zou2022} and computed using the FMMAX-backed implementation of the Bayer sorter challenge. The plotted efficiencies are the fraction of incident power transmitted into red, green, and blue subpixels.}
\label{figure:bayer_validation}
\end{figure}

\pfigref{figure:bayer_validation} plots the wavelength-dependent transmission into red, green, and blue subpixels for the Bayer color sorter design from \cite{Zou2022}. Transmission values obtained by Lumerical FDTD calculations (plotted in supplementary Fig. 7 of \cite{Zou2022}) and by the FMMAX-backed Bayer challenge implementation are shown to be in close agreement.

Table \ref{table:metalens} gives the intensity enhancement at the focus for eight metalens designs from \cite{imageruler} having variable nominal minimum length scale. The designs were generated in \cite{imageruler} using either FEM-backed or FDTD-backed topology optimization pipelines and then cross-validated. The intensity enhancement from FEM and FDTD calculations in \cite{imageruler} are given along with values from the FMMAX-backed implementation of the metalens challenge; all are found to be in good agreement.

\begin{table}[t]
\footnotesize
\begin{center}
\setlength{\tabcolsep}{4pt}
\begin{tabular}{ r | c c c | c c c | c c c }
  & \multicolumn{3}{c}{Comsol (FEM) \cite{imageruler}} &
  \multicolumn{3}{c}{Meep (FDTD) \cite{imageruler}} &
  \multicolumn{3}{c}{FMMAX (RCWA)} \\ 
 \hline
  & 450 nm & 550 nm & 650 nm & 450 nm & 550 nm & 650 nm & 450 nm & 550 nm & 650 nm \\
 \hline
Rasmus 70 nm & 21.1 & 20.2 & 21.2 & 21.8 & 23.7 & 24.2 & 20.0 & 21.2 & 22.8\\
Rasmus 123 nm & 16.3 & 15.2 & 16.6 & 16.3 & 14.9 & 15.0 & 16.5 & 14.5 & 15.6 \\
Rasmus 209 nm & 11.7 & 11.7 & 11.6 & 12.1 & 11.3 & 11.0 & 11.5 & 10.6 & 10.8\\
Rasmus 256 nm & 8.4 & 7.6 & 8.5 & 7.6 & 7.8 & 8.5 & 9.2 & 7.9 & 7.9 \\
Mo 86 nm & 16.1 & 14.9 & 16.3 & 14.9 & 15.3 & 16.7 & 14.5 & 15.4 & 16.9 \\
Mo 117 nm & 12.5 & 12.1 & 12.8 & 12.7 & 12.1 & 12.3 & 13.0 & 12.4 & 13.1 \\
Mo 180 nm & 12.5 & 12.0 & 12.0 & 12.8 & 12.1 & 12.6 & 12.1 & 12.1 & 13.6\\
Mo 242 nm & 10.9 & 11.4 & 11.6 & 10.8 & 11.4 & 11.5 & 10.1 & 11.7 & 11.9 \\
\end{tabular}
\end{center}
\caption{Figures of merit for metalens designs as reported in \cite{imageruler} and computed using the FMMAX-backed implementation of the metalens challenge. The \textit{Rasmus} designs were created using an FEM-backed topology optimization pipeline; the \textit{Mo} designs were created using an FDTD-backed pipeline.}
\label{table:metalens}
\end{table}

Table \ref{table:photon_extractor} gives the flux and DOS enhancement for $x$-, $y$-, and $z$-oriented point dipoles for the photon extractor design in \cite{Chakravarthi:20}. Values are given for the FDTD calculations in \cite{Chakravarthi:20} (from Fig. 1 (c) of that reference) and for RCWA calculations from the gym challenge. Due to the fundamental difficulty of representing point sources with the RCWA Fourier basis, a large number of Fourier terms $N$ is required to achieve convergence. While there is discrepancy between reference results and those for lower $N$, the low-$N$ the calculations still represent accurate solutions to Maxwell's equations, albeit for the case where the dipole source has finite spatial extent.

\begin{table}[t]
\footnotesize
\begin{center}
\setlength{\tabcolsep}{4pt}
\begin{tabular}{ r | c | c | c | c }
  & Meep (FDTD) \cite{Chakravarthi:20}
  & \multicolumn{3}{c}{FMMAX (RCWA)} \\
& & $N$=1200 & $N$=2400 & $N$=3000 \\
 \hline
$x$-dipole flux enhancement & 10.9 & 8.37 & 10.4 & 10.7 \\
$y$-dipole flux enhancement & 10.9 & 8.72 & 10.6 & 10.8\\
$z$-dipole flux enhancement & 352 & 182 & 257 & 274 \\
$x$-dipole LDOS enhancement & 1.41 & 1.31 & 1.27 & 1.29 \\
$y$-dipole LDOS enhancement & 1.41 & 1.32 & 1.28 & 1.30 \\
$z$-dipole LDOS enhancement & 1.35 & 1.26 & 1.34 & 1.36\\
\end{tabular}
\end{center}
\caption{Flux and LDOS enhancements of point dipoles for the photon extractor design from \cite{Chakravarthi:20}, as calculated using Meep (FDTD) in \cite{Chakravarthi:20} and using the FMMAX-backed implementation of the photon extractor challenge. All values are for the Nitrogen vacancy zero phonon line of 637 nm. The FMMAX values are given for various number of Fourier terms $N$; for lower $N$, the point dipole cannot be properly resolved, leading to a decrease in enhancement.
}
\label{table:photon_extractor}
\end{table}


\pagebreak
\section*{Appendix C: Compute cost for gym challenges}

Table \ref{table:speed_test} gives the step time for each challenge problem with default configuration and 32 bit precision. The step time is the time required to perform a simulation, evaluate the loss function, and compute the gradient of loss with respect to the design parameters. This is repeated five times, and the mean of the three fastest step times is reported.

The step time was measured on a workstation with a 28-core Intel Xeon w7-3465x processor, eight memory channels with 128 GB of RAM, and a single Nvidia RTX 4090 GPU with 24 GB of VRAM.

In general, default challenge configurations are intended for use in an optimization context--trading off some accuracy for faster iteration, e.g. by using a lower simulation resolution or fewer wavelengths. To further improve throughput, challenge problems can be "batched", i.e. the calculation described above can be performed in parallel for a batch of distinct design parameters. As shown in table \ref{table:speed_test}, in many cases this incurs only minimal additional step time. The maximum possible degree of parallelism is generally limited by GPU memory. Ceviche challenges can also be batched, but this is not found to yield performance benefits.

\begin{table}[ht!]
\footnotesize
\begin{center}
\begin{tabular}{ r | c | c | c | c}
& single & \multicolumn{3}{c}{batched} \\
 challenge & step time (s) & batch size & step time (s) & effective time (s) \\ 
 \hline
 metagrating & 0.31 & 20 & 0.70 &  0.035 \\  
 diffractive splitter & 4.91 & 10 & 6.91 & 0.691 \\  
 meta-atom library & 2.11 & 10 & 3.36 & 0.336\\  
 Bayer sorter & 5.93 & 10 & 9.28 & 0.928\\  
 metalens & 23.93 & 2 & 25.96 & 12.98 \\  
  Ceviche beam splitter & 9.76 & &  \\  
 Ceviche mode converter & 5.45 & & \\  
 Ceviche power splitter & 5.23 & & \\  
 Ceviche waveguide bend & 5.44 & & \\  
 Ceviche wdm & 44.59 & & \\  
 Ceviche lightweight beam splitter & 0.50 & & \\  
 Ceviche lightweight mode converter & 0.29 & & \\  
 Ceviche lightweight power splitter & 0.30 & & \\  
 Ceviche lightweight waveguide bend & 0.30 & & \\  
 Ceviche lightweight wdm & 0.51 & & \\
 photon extractor & 54.00 & 4 & 59.60 & 14.90
\end{tabular}
\end{center}
\caption{Step time for nanophotonic design challenges. The step time is the time required to compute the loss along with the gradient of loss with respect to the optimization variables. For some challenges, improved performance can be reached by running several instances in parallel. Tests were run on a workstation with a 28-core Intel Xeon w7-3465X processor, eight memory channels with 128 GB of RAM, and a single Nvidia RTX 4090 GPU with 24 GB of VRAM.}
\label{table:speed_test}
\end{table}


\pagebreak
\section*{Appendix D: Design with small measured length scale}
Several reference solutions to the challenges are measured to have small minimum length scale despite restrictions on the design space intended to promote manufacturable solutions. For example, diffractive splitter designs from \cite{LightTrans} uses a design grid with coarse 400 nm pixel size. The Bayer sorter design from \cite{Zou2022} uses a design grid with 125 nm pixel size. The photon extractor solution from \cite{Chakravarthi:20} used density and binarization filters to restrict minimum feature size to 50 nm. However, the measured minimum length scales for these are 80-360 nm, 20 nm, and 5 nm, respectively. This outcome is due to the details of the \code{imageruler} algorithm \cite{imageruler}, for which strong correspondence between larger measured length scales and reliable manufacturability is intended.

A brief description of the \code{imageruler} algorithm follows; for details, the reader is referred to \cite{imageruler} and the associated repository \cite{imageruler-repo}. The algorithm takes a binary array as input and tests whether solid and void features in the array can be realized with pixelated circular kernels of varying size. The minimum length scale for solid (void) features is the diameter of the largest kernel that can produce the solid (void) features in the array, with some small tolerance for violations at the edges of large features. The algorithm reports length scale in pixels, which in this work is converted to physical units by scaling with the pixel size of the design array.

\begin{figure}[ht!]
\centering\includegraphics{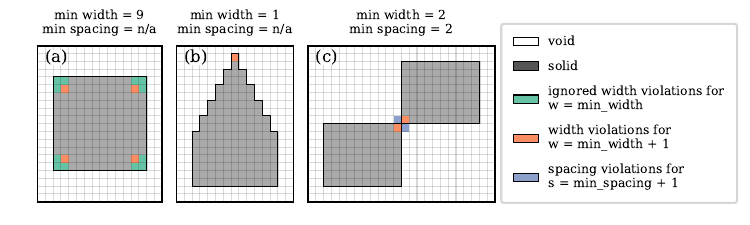}
\caption{Several arrays with small features, their measured length scales, and pixels that constitute violations. (a) A square solid feature that is 12$\times$12 pixels in size with measured minimum width of 9 pixels. By default, certain violations at the edges of large features are ignored; these are indicated in green for a test kernel that is 9 pixels wide. When test kernel is 10 pixels wide, additional interior pixels constitute violations. These are indicated in red and not ignored, hence the reported minimum width of 9. (b) An 11-pixel wide square feature that terminates with a sharp tip. The pixel at the tip constitutes a violation for a size-2 test kernel. (c) A checkerboard feature with measured minimum width and spacing of 2 pixels. Violations at the central corner pixels--unlike other corner pixels in (c) or the corners in (a)--are \textit{not} ignored by the \code{imageruler} algorithm.}
\label{figure:problematic_features}
\end{figure}

\pfigref{figure:problematic_features} depicts three example features for which small length scales are reported and are representative of features in the designs referenced above. Fig \ref{figure:problematic_features} (a) is a square fetaure that is 12 pixels in width; it is measured as having a minimum width of 9 pixels due to the inability of larger circular kernels to realize the square feature without missing non-edge pixels near the corners. (Strictly speaking, even a size-9 kernel yields corner violations, but this is allowed by tolerance within the \code{imageruler} algorithm.) \figref{figure:problematic_features} (b) is an 11-pixel wide square feature terminates with a sharp tip; it is measured as having a minimum width of 1 pixel. Finally, \figref{figure:problematic_features} (c) depicts two rectangular features touching at the corner, which is measured as having a minimum width and spacing both equal to 2 pixels. The algorithm does not ignore these violations as they are not considered to be at an edge. Features such as those in \figref{figure:problematic_features} are generally difficult to manufacture with consistency, with sharp features tending to round and checkerboard features randomly bridging or forming islands.

\pagebreak

\section*{Appendix E: Scattering parameters for Ceviche challenge solutions}
\begin{figure}[ht!]
\centering\includegraphics{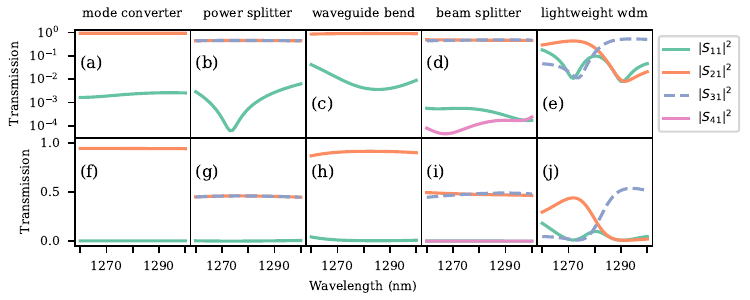}
\caption{Scattering parameters for Ceviche challenge solutions in \figref{figure:Ceviche}.}
\label{figure:Ceviche_sparams}
\end{figure}

\pfigref{figure:Ceviche_sparams} plots the scattering parameters for the Ceviche challenge solutions shown in \figref{figure:Ceviche}.

\begin{backmatter}

\bmsection{Acknowledgments}{I thank Aaditya Chandrasekhar and Rahul Kumar Padhy for contributing designs generated by the PhoTOS scheme, and Ian A. D. Williamson for contributing designs created by feasible generator and gradient estimator strategy.}

\bmsection{Disclosures}{The author declares no conflicts of interest.}

\bmsection{Data availability}{Data underlying the results presented in this paper are available in the leaderboard repository at \href{https://github.com/invrs-io/leaderboard}{github.com/invrs-io/leaderboard}.}

\bibliography{references}

\begin{thebibliography}{10}
\newcommand{\enquote}[1]{``#1''}

\bibitem{elesin2014time}
Y.~Elesin, B.~S. Lazarov, J.~S. Jensen, and O.~Sigmund, \enquote{Time domain topology optimization of 3d nanophotonic devices,} {\protect\JournalTitle{Photonics and Nanostructures-Fundamentals and Applications}} \textbf{12}, 23--33 (2014).

\bibitem{frellsen2016topology}
L.~F. Frellsen, Y.~Ding, O.~Sigmund, and L.~H. Frandsen, \enquote{Topology optimized mode multiplexing in silicon-on-insulator photonic wire waveguides,} {\protect\JournalTitle{Optics express}} \textbf{24}, 16866--16873 (2016).

\bibitem{Hammond:21}
A.~M. Hammond, A.~Oskooi, S.~G. Johnson, and S.~E. Ralph, \enquote{Photonic topology optimization with semiconductor-foundry design-rule constraints,} {\protect\JournalTitle{Opt. Express}} \textbf{29}, 23916--23938 (2021).

\bibitem{Hammond:22}
A.~M. Hammond, A.~Oskooi, M.~Chen, \emph{et~al.}, \enquote{High-performance hybrid time/frequency-domain topology optimization for large-scale photonics inverse design,} {\protect\JournalTitle{Opt. Express}} \textbf{30}, 4467--4491 (2022).

\bibitem{Vercruysse2019}
D.~Vercruysse, N.~V. Sapra, L.~Su, \emph{et~al.}, \enquote{Analytical level set fabrication constraints for inverse design,} {\protect\JournalTitle{Scientific Reports}} \textbf{9}, 8999 (2019).

\bibitem{piggott2017fabrication}
A.~Y. Piggott, J.~Petykiewicz, L.~Su, and J.~Vu{\v{c}}kovi{\'c}, \enquote{Fabrication-constrained nanophotonic inverse design,} {\protect\JournalTitle{Scientific reports}} \textbf{7}, 1786 (2017).

\bibitem{piggott2020inverse}
A.~Y. Piggott, E.~Y. Ma, L.~Su, \emph{et~al.}, \enquote{Inverse-designed photonics for semiconductor foundries,} {\protect\JournalTitle{Acs Photonics}} \textbf{7}, 569--575 (2020).

\bibitem{shen2015integrated}
B.~Shen, P.~Wang, R.~Polson, and R.~Menon, \enquote{An integrated-nanophotonics polarization beamsplitter with 2.4$\times$ 2.4 $\mu$m2 footprint,} {\protect\JournalTitle{Nature Photonics}} \textbf{9}, 378--382 (2015).

\bibitem{Tahersima2019}
M.~H. Tahersima, K.~Kojima, T.~Koike-Akino, \emph{et~al.}, \enquote{Deep neural network inverse design of integrated photonic power splitters,} {\protect\JournalTitle{Scientific Reports}} \textbf{9}, 1368 (2019).

\bibitem{GOUDARZI2022105268}
K.~Goudarzi and M.~Lee, \enquote{Inverse design of a binary waveguide crossing by the particle swarm optimization algorithm,} {\protect\JournalTitle{Results in Physics}} \textbf{34}, 105268 (2022).

\bibitem{Zhaocheng}
Z.~Liu, D.~Zhu, L.~Raju, and W.~Cai, \enquote{Tackling photonic inverse design with machine learning,} {\protect\JournalTitle{Advanced Science}} \textbf{8}, 2002923 (2021).

\bibitem{Jiang2021}
J.~Jiang, M.~Chen, and J.~A. Fan, \enquote{Deep neural networks for the evaluation and design of photonic devices,} {\protect\JournalTitle{Nature Reviews Materials}} \textbf{6}, 679--700 (2021).

\bibitem{Wiecha:21}
P.~R. Wiecha, A.~Arbouet, C.~Girard, and O.~L. Muskens, \enquote{Deep learning in nano-photonics: inverse design and beyond,} {\protect\JournalTitle{Photon. Res.}} \textbf{9}, B182--B200 (2021).

\bibitem{imageruler}
M.~Chen, R.~E. Christiansen, J.~A. Fan, \emph{et~al.}, \enquote{Validation and characterization of algorithms and software for photonics inverse design,} {\protect\JournalTitle{J. Opt. Soc. Am. B}} \textbf{41}, A161--A176 (2024).

\bibitem{Schubert2022}
M.~F. Schubert, A.~K.~C. Cheung, I.~A.~D. Williamson, \emph{et~al.}, \enquote{Inverse design of photonic devices with strict foundry fabrication constraints,} {\protect\JournalTitle{ACS Photonics}} \textbf{9}, 2327--2336 (2022).

\bibitem{common_task_method}
D.~Donoho, \enquote{50 years of data science,} {\protect\JournalTitle{Journal of Computational and Graphical Statistics}} \textbf{26}, 745--766 (2017).

\bibitem{deng2009imagenet}
J.~Deng, W.~Dong, R.~Socher, \emph{et~al.}, \enquote{Imagenet: A large-scale hierarchical image database,} in \emph{2009 IEEE conference on computer vision and pattern recognition,}  (Ieee, 2009), pp. 248--255.

\bibitem{lecun1998gradient}
Y.~LeCun, L.~Bottou, Y.~Bengio, and P.~Haffner, \enquote{Gradient-based learning applied to document recognition,} {\protect\JournalTitle{Proceedings of the IEEE}} \textbf{86}, 2278--2324 (1998).

\bibitem{krizhevsky2009learning}
A.~Krizhevsky, G.~Hinton \emph{et~al.}, \enquote{Learning multiple layers of features from tiny images,}  (2009).

\bibitem{posebusters}
M.~Buttenschoen, G.~M. Morris, and C.~M. Deane, \enquote{Posebusters: Ai-based docking methods fail to generate physically valid poses or generalise to novel sequences,} {\protect\JournalTitle{Chemical Science}} \textbf{15}, 3130--3139 (2024).

\bibitem{basu2016dockq}
S.~Basu and B.~Wallner, \enquote{Dockq: a quality measure for protein-protein docking models,} {\protect\JournalTitle{PloS one}} \textbf{11}, e0161879 (2016).

\bibitem{common_task_method_social_science}
M.~J. Salganik, I.~Lundberg, A.~T. Kindel, \emph{et~al.}, \enquote{Measuring the predictability of life outcomes with a scientific mass collaboration,} {\protect\JournalTitle{Proceedings of the National Academy of Sciences}} \textbf{117}, 8398--8403 (2020).

\bibitem{Angeris:21}
G.~Angeris, J.~Vu\v{c}kovi\'{c}, and S.~Boyd, \enquote{Heuristic methods and performance bounds for photonic design,} {\protect\JournalTitle{Opt. Express}} \textbf{29}, 2827--2854 (2021).

\bibitem{sigmund2022benchmarking}
O.~Sigmund, \enquote{On benchmarking and good scientific practise in topology optimization,} {\protect\JournalTitle{Structural and Multidisciplinary Optimization}} \textbf{65}, 315 (2022).

\bibitem{CHRISTIANSEN201923}
R.~E. Christiansen, J.~Vester-Petersen, S.~P. Madsen, and O.~Sigmund, \enquote{A non-linear material interpolation for design of metallic nano-particles using topology optimization,} {\protect\JournalTitle{Computer Methods in Applied Mechanics and Engineering}} \textbf{343}, 23--39 (2019).

\bibitem{totypes-repo}
M.~F. Schubert, \enquote{totypes: Custom types for topology optimization,} \url{https://github.com/invrs-io/totypes} (2023).

\bibitem{filter-project}
F.~Wang, B.~S. Lazarov, and O.~Sigmund, \enquote{On projection methods, convergence and robust formulations in topology optimization,} {\protect\JournalTitle{Structural and Multidisciplinary Optimization}} \textbf{43}, 767--784 (2011).

\bibitem{hazineh2022dflat}
D.~S. Hazineh, S.~W.~D. Lim, Z.~Shi, \emph{et~al.}, \enquote{D-flat: A differentiable flat-optics framework for end-to-end metasurface visual sensor design,}  (2022).

\bibitem{Yeung2023}
C.~Yeung, B.~Pham, R.~Tsai, \emph{et~al.}, \enquote{Deepadjoint: An all-in-one photonic inverse design framework integrating data-driven machine learning with optimization algorithms,} {\protect\JournalTitle{ACS Photonics}} \textbf{10}, 884--891 (2023).

\bibitem{no_free_lunch}
D.~Wolpert and W.~Macready, \enquote{No free lunch theorems for optimization,} {\protect\JournalTitle{IEEE Transactions on Evolutionary Computation}} \textbf{1}, 67--82 (1997).

\bibitem{Sell2017}
D.~Sell, J.~Yang, S.~Doshay, \emph{et~al.}, \enquote{Large-angle, multifunctional metagratings based on freeform multimode geometries,} {\protect\JournalTitle{Nano Letters}} \textbf{17}, 3752--3757 (2017).

\bibitem{metagrating-01}
J.~Yang, D.~Sell, and J.~A. Fan, \enquote{Freeform metagratings based on complex light scattering dynamics for extreme, high efficiency beam steering,} {\protect\JournalTitle{Annalen der Physik}} \textbf{530}, 1700302 (2018).

\bibitem{Sabzevari:24}
A.~Sabzevari and A.~Hatef, \enquote{Inverse design and optimization of a one-dimensional metagrating beam deflector by smart pattern search,} {\protect\JournalTitle{Appl. Opt.}} \textbf{63}, 4793--4798 (2024).

\bibitem{Ye:22}
T.~Ye, D.~Wu, Q.~Wu, \emph{et~al.}, \enquote{Realization of inversely designed metagrating for highly efficient large angle beam deflection,} {\protect\JournalTitle{Opt. Express}} \textbf{30}, 7566--7579 (2022).

\bibitem{Chen:21}
W.-Q. Chen, D.-S. Zhang, S.-Y. Long, \emph{et~al.}, \enquote{Nearly dispersionless multicolor metasurface beam deflector for near eye display designed by a physics-driven deep neural network,} {\protect\JournalTitle{Appl. Opt.}} \textbf{60}, 3947--3953 (2021).

\bibitem{Xiao:24}
Y.~Xiao, M.~Xu, M.~Pu, \emph{et~al.}, \enquote{Topology-optimized freeform broadband optical metagrating for high-efficiency large-angle deflection,} {\protect\JournalTitle{J. Opt. Soc. Am. B}} \textbf{41}, A52--A59 (2024).

\bibitem{Jenkins2023}
R.~P. Jenkins, S.~D. Campbell, and D.~H. Werner, \enquote{General-purpose algorithm for two-material minimum feature size enforcement of freeform nanophotonic devices,} {\protect\JournalTitle{ACS Photonics}} \textbf{10}, 845--853 (2023).

\bibitem{fmmax}
M.~F. Schubert and A.~M. Hammond, \enquote{Fourier modal method for inverse design of metasurface-enhanced micro-leds,} {\protect\JournalTitle{Opt. Express}} \textbf{31}, 42945--42960 (2023).

\bibitem{meep}
A.~F. Oskooi, D.~Roundy, M.~Ibanescu, \emph{et~al.}, \enquote{{MEEP}: A flexible free-software package for electromagnetic simulations by the {FDTD} method,} {\protect\JournalTitle{Computer Physics Communications}} \textbf{181}, 687--702 (2010).

\bibitem{hugonin2021reticolo}
J.~P. Hugonin and P.~Lalanne, \enquote{Reticolo software for grating analysis,} {\protect\JournalTitle{arXiv preprint arXiv:2101.00901}}  (2021).

\bibitem{structured_light_1}
R.~{Vandenhouten}, A.~{Hermerschmidt}, and R.~{Fiebelkorn}, \enquote{{Design and quality metrics of point patterns for coded structured light illumination with diffractive optical elements in optical 3D sensors},} in \emph{Society of Photo-Optical Instrumentation Engineers (SPIE) Conference Series,}  vol. 10335 of \emph{Society of Photo-Optical Instrumentation Engineers (SPIE) Conference Series} B.~C. {Kress}, W.~{Osten}, and H.~P. {Urbach}, eds. (2017), p. 1033518.

\bibitem{structured_light_2}
O.~Barlev and M.~A. Golub, \enquote{Multifunctional binary diffractive optical elements for structured light projectors,} {\protect\JournalTitle{Opt. Express}} \textbf{26}, 21092--21107 (2018).

\bibitem{multifocal_1}
J.~E. Jureller, H.~Y. Kim, and N.~F. Scherer, \enquote{Stochastic scanning multiphoton multifocal microscopy,} {\protect\JournalTitle{Opt. Express}} \textbf{14}, 3406--3414 (2006).

\bibitem{multifocal_2}
Z.~Chen, B.~Mc~Larney, J.~Rebling, \emph{et~al.}, \enquote{High-speed large-field multifocal illumination fluorescence microscopy,} {\protect\JournalTitle{Laser \& Photonics Reviews}} \textbf{14}, 1900070 (2020).

\bibitem{LightTrans}
LightTrans, \enquote{Design and rigorous analysis of non-paraxial diffractive beam splitter,} \url{https://www.lighttrans.com/use-cases/application/design-and-rigorous-analysis-of-non-paraxial-diffractive-beam-splitter.html}. Version: 3.1.

\bibitem{Kim:20}
D.~C. Kim, A.~Hermerschmidt, P.~Dyachenko, and T.~Scharf, \enquote{Inverse design and demonstration of high-performance wide-angle diffractive optical elements,} {\protect\JournalTitle{Opt. Express}} \textbf{28}, 22321--22333 (2020).

\bibitem{pestourie2018inverse}
R.~Pestourie, C.~P{\'e}rez-Arancibia, Z.~Lin, \emph{et~al.}, \enquote{Inverse design of large-area metasurfaces,} {\protect\JournalTitle{Optics express}} \textbf{26}, 33732--33747 (2018).

\bibitem{li2022inverse}
Z.~Li, R.~Pestourie, J.-S. Park, \emph{et~al.}, \enquote{Inverse design enables large-scale high-performance meta-optics reshaping virtual reality,} {\protect\JournalTitle{Nature communications}} \textbf{13}, 1--11 (2022).

\bibitem{Chen2023}
W.~T. Chen, J.-S. Park, J.~Marchioni, \emph{et~al.}, \enquote{Dispersion-engineered metasurfaces reaching broadband 90{\%} relative diffraction efficiency,} {\protect\JournalTitle{Nature Communications}} \textbf{14}, 2544 (2023).

\bibitem{phan2019high}
T.~Phan, D.~Sell, E.~W. Wang, \emph{et~al.}, \enquote{High-efficiency, large-area, topology-optimized metasurfaces,} {\protect\JournalTitle{Light: Science \& Applications}} \textbf{8}, 48 (2019).

\bibitem{zhou2024large}
Y.~Zhou, C.~Mao, E.~Gershnabel, \emph{et~al.}, \enquote{Large-area, high-numerical-aperture, freeform metasurfaces,} {\protect\JournalTitle{Laser \& Photonics Reviews}} p. 2300988 (2024).

\bibitem{Zou2022}
X.~Zou, Y.~Zhang, R.~Lin, \emph{et~al.}, \enquote{Pixel-level bayer-type colour router based on metasurfaces,} {\protect\JournalTitle{Nature Communications}} \textbf{13}, 3288 (2022).

\bibitem{Peng:23}
Y.~Q. Peng, H.~P. Lu, D.~S. Zhang, \emph{et~al.}, \enquote{Inverse design of a light nanorouter for a spatially multiplexed optical filter,} {\protect\JournalTitle{Opt. Lett.}} \textbf{48}, 6232--6235 (2023).

\bibitem{Miyata:21}
M.~Miyata, N.~Nemoto, K.~Shikama, \emph{et~al.}, \enquote{Full-color-sorting metalenses for high-sensitivity image sensors,} {\protect\JournalTitle{Optica}} \textbf{8}, 1596--1604 (2021).

\bibitem{Chen2017}
B.~H. Chen, P.~C. Wu, V.-C. Su, \emph{et~al.}, \enquote{Gan metalens for pixel-level full-color routing at visible light,} {\protect\JournalTitle{Nano Letters}} \textbf{17}, 6345--6352 (2017).

\bibitem{submicron_bayer}
S.~Yun, S.~Roh, S.~Lee, \emph{et~al.}, \enquote{Highly efficient color separation and focusing in the sub-micron cmos image sensor,} in \emph{2021 IEEE International Electron Devices Meeting (IEDM),}  (2021), pp. 30.1.1--30.1.4.

\bibitem{CatrysseZhaoJinFan}
P.~B. Catrysse, N.~Zhao, W.~Jin, and S.~Fan, \enquote{Subwavelength bayer rgb color routers with perfect optical efficiency,} {\protect\JournalTitle{Nanophotonics}} \textbf{11}, 2381--2387 (2022).

\bibitem{freeform_submicron}
C.~Kim, J.~Hong, J.~Jang, \emph{et~al.}, \enquote{Freeform metasurface color router for deep submicron pixel image sensors,} {\protect\JournalTitle{Science Advances}} \textbf{10}, eadn9000 (2024).

\bibitem{VIS/NIR1}
N.~Zhao, P.~B. Catrysse, and S.~Fan, \enquote{Perfect rgb-ir color routers for sub-wavelength size cmos image sensor pixels,} {\protect\JournalTitle{Advanced Photonics Research}} \textbf{2}, 2000048 (2021).

\bibitem{VIS/NIR2}
R.~Zhong, X.~Xu, Y.~Zhou, \emph{et~al.}, \enquote{High-efficiency integrated color routers by simple identical nanostructures for visible and near-infrared wavelengths,} {\protect\JournalTitle{Photonics}} \textbf{10} (2023).

\bibitem{VIS/NIR3}
Y.~Shao, S.~Guo, R.~Chen, \emph{et~al.}, \enquote{Pixelated nir–vis spectral routers based on 2d mie-type metagratings,} {\protect\JournalTitle{Laser \& Photonics Reviews}} \textbf{17}, 2300027 (2023).

\bibitem{VIS/NIR4}
Y.~J. Hong, B.~J. Jeon, Y.~G. Ki, and S.~J. Kim, \enquote{A metasurface color router facilitating rgb-nir sensing for an image sensor application,} {\protect\JournalTitle{Nanophotonics}} \textbf{13}, 1407--1415 (2024).

\bibitem{ZOU2023107472}
X.~Zou, G.~Gong, Y.~Lin, \emph{et~al.}, \enquote{Metasurface-based polarization color routers,} {\protect\JournalTitle{Optics and Lasers in Engineering}} \textbf{163}, 107472 (2023).

\bibitem{Christiansen:21}
R.~E. Christiansen and O.~Sigmund, \enquote{Inverse design in photonics by topology optimization: tutorial,} {\protect\JournalTitle{J. Opt. Soc. Am. B}} \textbf{38}, 496--509 (2021).

\bibitem{Li2022}
Z.~Li, R.~Pestourie, Z.~Lin, \emph{et~al.}, \enquote{Empowering metasurfaces with inverse design: Principles and applications,} {\protect\JournalTitle{ACS Photonics}} \textbf{9}, 2178--2192 (2022).

\bibitem{metalens2}
M.~Khorasaninejad, W.~T. Chen, R.~C. Devlin, \emph{et~al.}, \enquote{Metalenses at visible wavelengths: Diffraction-limited focusing and subwavelength resolution imaging,} {\protect\JournalTitle{Science}} \textbf{352}, 1190--1194 (2016).

\bibitem{metalens3}
Z.~Lin, B.~Groever, F.~Capasso, \emph{et~al.}, \enquote{Topology-optimized multilayered metaoptics,} {\protect\JournalTitle{Phys. Rev. Appl.}} \textbf{9}, 044030 (2018).

\bibitem{Christiansen:20}
R.~E. Christiansen, Z.~Lin, C.~Roques-Carmes, \emph{et~al.}, \enquote{Fullwave maxwell inverse design of axisymmetric, tunable, and multi-scale multi-wavelength metalenses,} {\protect\JournalTitle{Opt. Express}} \textbf{28}, 33854--33868 (2020).

\bibitem{Chung:20}
H.~Chung and O.~D. Miller, \enquote{High-na achromatic metalenses by inverse design,} {\protect\JournalTitle{Opt. Express}} \textbf{28}, 6945--6965 (2020).

\bibitem{ceviche-challenges}
I.~A.~D. Williamson, \enquote{Ceviche challenges: photonic inverse design suite,} \url{https://github.com/google/ceviche-challenges} (2022).

\bibitem{Hughes2019}
T.~W. Hughes, I.~A.~D. Williamson, M.~Minkov, and S.~Fan, \enquote{Forward-mode differentiation of maxwell's equations,} {\protect\JournalTitle{ACS Photonics}} \textbf{6}, 3010--3016 (2019).

\bibitem{surco}
A.~M. Ferber, T.~Huang, D.~Zha, \emph{et~al.}, \enquote{{S}ur{C}o: Learning linear {SUR}rogates for {CO}mbinatorial nonlinear optimization problems,} in \emph{Proceedings of the 40th International Conference on Machine Learning,}  vol. 202 of \emph{Proceedings of Machine Learning Research} A.~Krause, E.~Brunskill, K.~Cho, \emph{et~al.}, eds. (PMLR, 2023), pp. 10034--10052.

\bibitem{padhy2024PhoTOS}
R.~K. Padhy and A.~Chandrasekhar, \enquote{Photos: Topology optimization of photonic components using a shape library,}  (2024).

\bibitem{ferber2023genco}
A.~Ferber, A.~Zharmagambetov, T.~Huang, \emph{et~al.}, \enquote{Genco: Generating diverse solutions to design problems with combinatorial nature,} {\protect\JournalTitle{arXiv preprint arXiv:2310.02442}}  (2023).

\bibitem{Cheung:24}
A.~K.~C. Cheung, K.~Gadepalli, J.~Guan, \emph{et~al.}, \enquote{Inverse-designed cwdm demultiplexer operated in o-band,} in \emph{Optical Fiber Communication Conference (OFC) 2024,}  (Optica Publishing Group, 2024), p. W1A.6.

\bibitem{Chakravarthi:20}
S.~Chakravarthi, P.~Chao, C.~Pederson, \emph{et~al.}, \enquote{Inverse-designed photon extractors for optically addressable defect qubits,} {\protect\JournalTitle{Optica}} \textbf{7}, 1805--1811 (2020).

\bibitem{wang2018maximizing}
F.~Wang, R.~E. Christiansen, Y.~Yu, \emph{et~al.}, \enquote{Maximizing the quality factor to mode volume ratio for ultra-small photonic crystal cavities,} {\protect\JournalTitle{Applied Physics Letters}} \textbf{113} (2018).

\bibitem{icsiklar2022trade}
G.~I{\c{s}}iklar, P.~T. Kristensen, J.~M{\o}rk, \emph{et~al.}, \enquote{On the trade-off between mode volume and quality factor in dielectric nanocavities optimized for purcell enhancement,} {\protect\JournalTitle{Optics Express}} \textbf{30}, 47304--47314 (2022).

\bibitem{miller2016fundamental}
O.~D. Miller, A.~G. Polimeridis, M.~Homer~Reid, \emph{et~al.}, \enquote{Fundamental limits to optical response in absorptive systems,} {\protect\JournalTitle{Optics express}} \textbf{24}, 3329--3364 (2016).

\bibitem{liang2013formulation}
X.~Liang and S.~G. Johnson, \enquote{Formulation for scalable optimization of microcavities via the frequency-averaged local density of states,} {\protect\JournalTitle{Optics express}} \textbf{21}, 30812--30841 (2013).

\bibitem{imageruler-repo}
S.~G. Johnson, W.~Ma, A.~Oskooi, \emph{et~al.}, \enquote{Imageruler: Measure minimum solid/void length scales in binary images,} \url{https://github.com/NanoComp/imageruler} (2024).

\end{thebibliography}

\end{backmatter}

\end{document}